# A 3-Dimensional Simplex Modulation Format with Improved OSNR Performance Compared to DP-BPSK

Annika Dochhan[1], Helmut Grießer[2], Brian Teipen[3], Michael Eiselt[1]
(1) ADVA Optical Networking SE, Märzenquelle 1-3, 98617 Meiningen, Germany
(2) ADVA Optical Networking SE, Fraunhoferstr. 9a, 82152 Martinsried / Munich, Germany
(3) ADVA Optical Networking SE, 5755 Peachtree Industrial Blvd., Norcross GA 30092, USA
adochhan@advaoptical.com

## Kurzfassung

Das neuartige 3-dimensionale Modulationsformat 3D-Simplex bietet das Potenzial einer 1.2 dB höheren OSNR-Toleranz als DP-BPSK bei gleicher spektraler Breite, mit der Modulation von zwei Bits pro Symbol. Dieser Vorteil von 3D-Simplex wird hier experimentell verifiziert und die Leistungsfähigkeit bei nichtlinearer Übertragung evaluiert. Alle experimentellen Ergebnisse werden durch ausführliche Simulationen gestützt. Der Gewinn durch das 3D-Simplex-Format im stark nichtlinearen Bereich bleibt nicht erhalten, jedoch ist die Leistungsfähigkeit mit der von DP-DPSK vergleichbar.

## Abstract

The novel 3-dimensional modulation format 3D-Simplex offers potentially 1.2 dB higher OSNR tolerance than DP-DPSK while exhibiting the same spectral occupancy, modulating two bits per symbol. We verify this benefit experimentally and evaluate the transmission performance in a non-linear environment. All experimental results are confirmed by simulations. The benefit of 3D-Simplex is not maintained in the highly non-linear regime, but the performance is still comparable to that of DP-DPSK.

## 1 Introduction

Besides a low noise transmission line with only few distortions, optical transmission over very long distances is enabled by modulation formats which exhibit high tolerance towards optical noise. Recently, 4-dimensional modulation formats have been proposed, in which both signal polarizations are modulated interdependently to increase the OSNR tolerance [1]. Previously, we proposed a 3-dimensional Simplex format [2] in which two bits are encoded in one symbol, thus exhibiting the same spectral occupancy as DP-BPSK, but with higher robustness towards optical noise. The OSNR benefit in comparison to DP-BPSK is verified by experiments for the back-to-back case. Transmission over a single span of 300 km standard single-mode fiber with counter-directional Raman pumping is used to evaluate the performance in non-linear transmission environment. In this paper, the experimental setup of [2] is transferred into a simulation environment to support the previous results with numerical investigations.

## 2 The 3-dimensional Modulation Format 3D-Simplex

The complex envelope of an optical signal can be modulated in four dimensions, the I- and Q-phases and in each of the two orthogonal polarizations. Conventionally, for DP-BPSK or DP-QPSK formats, each of these dimensions is modulated independently, resulting in the transmission of one bit per dimension or 2 or 4 bits per symbol, respectively. For low error rates, the noise tolerance of modulation formats is determined by the minimum Cartesian distance of their constellation points. In a 3- or 4-dimensional format, the orthogonal dimensions are modulated interdependently to increase the OSNR tolerance.

A geometrical representation for the novel 3D-Simplex modulation format is shown in **Fig. 1**, together with the two dimensional representation of DP-BPSK. Constellation diagrams are given in **Fig. 2**. For 3D-Simplex, the x-polarization carries a QPSK constellation, while the y-polarization is BPSK modulated. The choice of the BPSK symbol is dependent on the currently transmitted QPSK symbol, as indicated by the different colors of the constellation points in **Fig. 2**. As indicated by **Fig. 1** the Cartesian distance $D_{min}$ of $\sqrt{8}$ for 3D-Simplex exceeds that of DP-BPSK, which is equal to 2. In addition, 3D-Simplex offers an average power $P_{avg}=3$ in contrast to DP-BPSK with $P_{avg}=2$ (see **Fig. 2**). Comparing two modulation formats with the same bandwidth, the difference in required OSNR can be given by determining $P_{avg}/D^2_{min}$ for both formats and calculating their ratio [3]. This is only valid for low error rates. With the given values, the improved OSNR tolerance for 3D-Simplex compared to DP-BPSK is 1.2 dB.

**Tab. 1** shows a possible the bit encoding for the 3D-Simplex modulation format. For the x-polarization two bits are encoded into one QPSK symbol, while the y-polarization is BPSK modulated with the XOR of these two bits. The main advantage of this representation is that it allows unbiased binary driving signals.

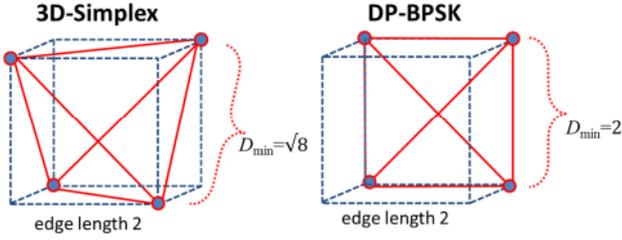

**Figure 1** Visualization of 3D-Simplex and DP-BPSK in a cube. The corners of the cube represent the constellation points.

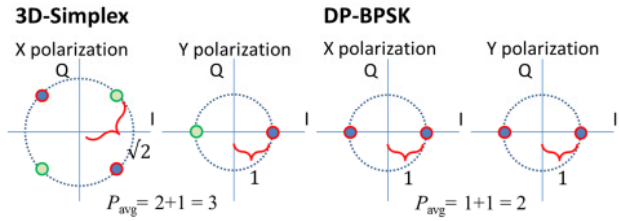

**Figure 2** Constellations of 3D-Simplex (left two diagrams) and DP-BPSK (right two diagrams) in the complex plane.

| Bit encoding | Ix | Qx | Iy | Qy |
|---|---|---|---|---|
| 00 | -1 | -1 | -1 | 0 |
| 01 | -1 | 1 | 1 | 0 |
| 10 | 1 | -1 | 1 | 0 |
| 11 | 1 | 1 | -1 | 0 |

**Table 1** Bit encoding of 3D-Simplex constellation points.

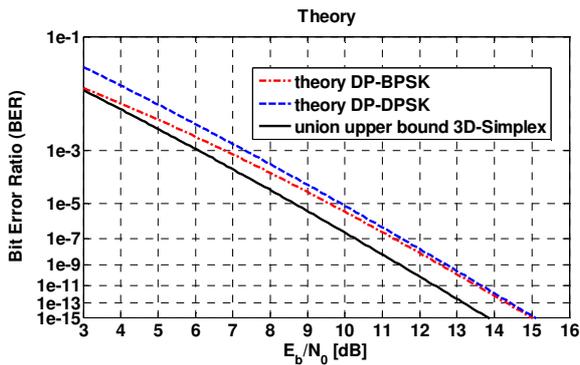

**Figure 3** Theoretical BER vs. $E_b/N_0$ for DP-BPSK, DP-DPSK and 3D-Simplex. For 3D-Simplex a union upper bound of the BER instead of an analytical result is given.

The 3D-Simplex format is not invariant for certain phase rotations and therefore does not require differential encoding. However, especially in highly non-linear regime, phase rotations will complicate the detection of DP-BPSK, thus the use of differentially pre-coded DP-DPSK might be advantageous. In the simulations and measurements we use DP-DPSK as comparison. Asymptotically, the performance of DP-BPSK equals that of DP-DPSK. However, in the BER range for systems using FEC, DP-BPSK outperforms DP-DPSK by several tenths of dB, as can be seen from **Fig. 3**, which shows the theoretical BER curves for DP-BPSK, DP-DPSK and a union upper bound for the BER of 3D-Simplex. To be independent of the data rate, the BER is displayed vs. bit energy $E_b$ over noise power spectrum density $N_0$.

## 3 System Setup

The 3D-Simplex modulation format was first tested in a back-to-back configuration for 16 and 25 GBaud. At 16 GBaud, it can be ensured that bandwidth limitations of the utilized components (namely the digital-to-analog converter (DAC) at the TX and the real time oscilloscope at the RX) do not significantly influence the system performance. In contrast, 25 GBaud is a possible line rate for 40 Gb/s long-haul transmission with sufficient FEC overhead. At 16 GBaud, single span transmission over 300 km demonstrates the formats' robustness towards non-linearity.

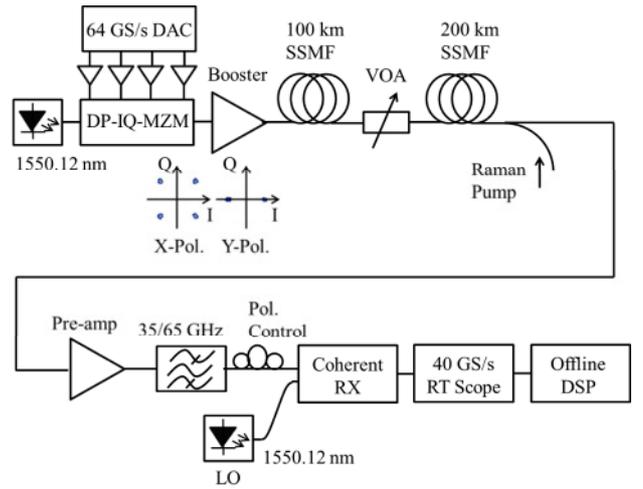

**Figure 4** Experimental setup. A DP-QPSK Mach-Zehnder-Modulator is used to generate a QPSK signal in x-polarization and a BPSK signal in the y-polarization. The signal is transmitted over a single 300 km span of SSMF using backward Raman pumping.

### 3.1 Experiment

The experimental setup is shown in **Fig. 4**. A DP-IQ-Mach-Zehnder modulator (DP-IQ-MZM) was driven by offline generated binary signals, which are stored inside the memory of a 64 GS/s DAC with 13 GHz electrical bandwidth [4]. This bandwidth limitation was partly mitigated be pre-compensation, using measured DAC transfer characteristics. The x-polarization was QPSK modulated with a $2^{11}$ de Bruijn sequence, while the I-branch of the y-polarization carried a BPSK signal, resulting from the XOR combination of the two QPSK encoded bits, according to **Tab. 1**. The Q-branch modulator of the y-polarization was biased at zero transmission. The Tx channel launch power into the fiber was varied between 10 and 20 dBm using a booster EDFA after the modulator. The transmission link consisted of 300 km SSMF with an attenuation of 63 dB. To increase the span loss, a variable optical attenuator (VOA) was placed after 100 km. Amplification was performed by 29 dBm counter-directional 4-wavelength Raman pumping from the receiving end. After transmission the 16 GBaud signal was filtered by a 35 GHz optical bandpass filter with flat-top shape, while for the 25 GBaud signal a 65 GHz filter was used. The filters emulate the demultiplexer in a 50 GHz or 100 GHz grid WDM transmission. The signal was detected with a standard coherent receiver, consisting of a local oscillator laser, a polarization beam splitter, two 90° hybrids and four balanced receivers. The linewidths of both lasers, at transmit and receive side, were approximately 100 kHz. The four electrical output signals of the balanced receivers were captured and stored by a real time (RT) oscilloscope exhibiting 40 GS/s and a bandwidth of 16 GHz. The stored data was processed offline in Python programming routines.

### 3.2 Simulations

The simulative investigations were performed in MATLAB$^{TM}$ and Python, using the same data sequence and pulse shape as for the measurement. The bandwidth of the DAC after pre-compensation and the limited bandwidth due to the oscilloscope at the Rx were taken into account by including measured transfer functions into the filter implementation of the simulations. BER results were obtained by Monte-Carlo simulations.

### 3.3 Signal Processing

Reception of the 3D-Simplex signal requires a coarse polarization pre-alignment, which is done in our case manually by a polarization controller (see **Fig. 3**). This is necessary since all parts of the DSP work blindly. For later implementations, the use of dedicated training symbols will eliminate the need for manual adjustment.

The offline DSP applies electrical matched filtering, followed by frequency domain chromatic dispersion compensation. The clock recovery uses a Gardner phase detector [5]. The 13 tap weights of the butterfly structure FIR filter for polarization separation are adapted blindly, using the observation of two consecutive output symbols for DPSK, as described in [6]. Since the 3D-Simplex is a hybrid of BPSK and QPSK, a combined equalization is applied to the tap weights for that format, using the algorithm of [6] and a standard CMA [7]. The frequency offset between the LO and the transmit laser is determined by the Mengali-Morelli algorithm [8], applied to the BPSK part of the signal. The Viterbi and Viterbi phase estimation [9] also works on the BPSK tributary and is amended by a phase offset estimation between the BPSK and the QPSK part. Especially in the case of non-linear transmission, this routine partially compensates for the increased phase rotation of the QPSK signal compared to BPSK.

In order to estimate the achievable performance of the new format, an idealized simulation with the standard CMA was performed that did not consider any polarization distortions or rotations. Moreover, ideal lasers were assumed, neglecting laser phase noise and frequency offset between transmit laser and LO.

## 4 Results

**Fig. 5** shows received back-to-back constellations of the 3D-Simplex format at an OSNR of 15.9 dB for experiments and simulations. Within the experiments, all received QPSK constellations include one slightly displaced constellation point as can be seen in the figure (right lower corner). One reason might be that the DP-IQ-MZM introduces cross talk when not driven by similar electrical signals on both polarizations.

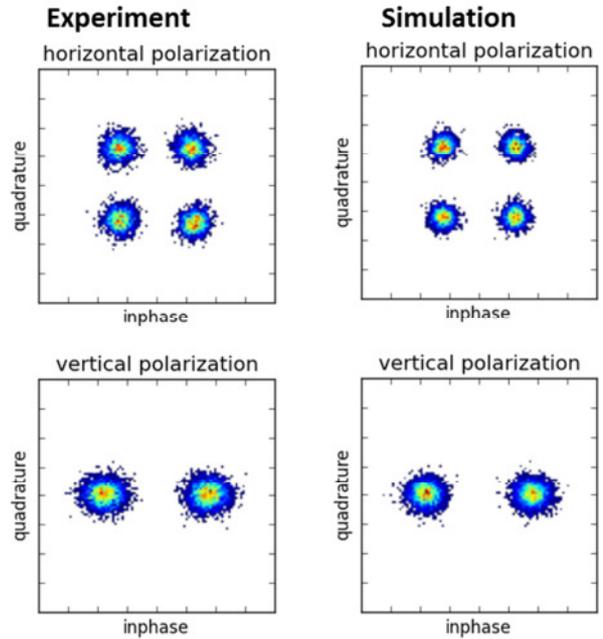

**Figure 5** Received constellations with an OSNR of 15.9 dB after DSP for experiment and simulation.

### 4.1 Experiments

Experimental and numerical BER results, obtained in a back-to-back configuration, are displayed in **Fig. 6** for 16 GBaud and in **Fig. 7** for 25 GBaud in comparison to theoretically expected values. The experiments show an implementation penalty of about 1 dB at 16 GBaud and 2-3 dB at 25 GBaud. The simulations indicate that ~0.3 dB at 16 GBaud and ~0.5 dB of this penalty result from narrow-band filtering. However, to explain the higher implementation penalty of 25 GBaud experiments it must be taken into account that the DAC pre-compensation might not work perfectly, since the Tx drivers are limiting amplifiers and in addition imply non-ideal flat transmission characteristics over the whole signal bandwidth. As predicted by theory, the 3D-Simplex requires about 1 dB less OSNR for the same BER as compared to DP-DPSK. Even as the implementation penalty increases for the higher symbol rate within the experiments, this benefit of 3D-Simplex is maintained. In all plots, the OSNR is defined with a 0.1 nm noise bandwidth, while the signal power was measured in a bandwidth of 0.5 nm.

Single channel transmission over 300 km SSMF was performed for the 16 GBaud signal, using variable input power. **Fig. 8** shows the resulting BER vs. channel launch power for both modulation formats in experiment (upper figure) and simulation (lower figure).

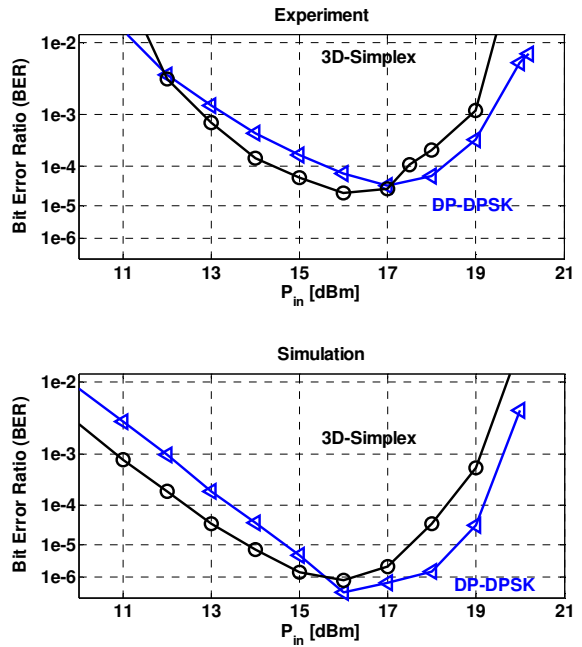

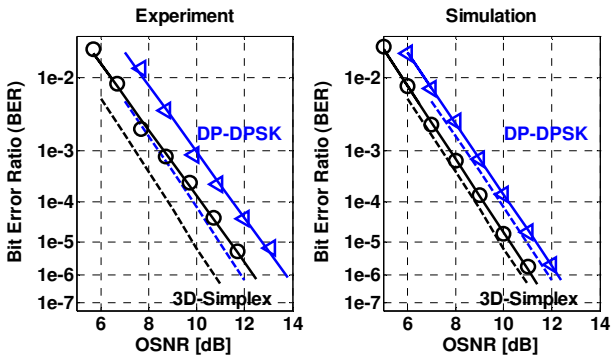

**Figure 6** Back-to-back BER results for 16 GBaud. 3D-Simplex (circles) and DP-DPSK (triangles) with solid lines as linear regression in comparison to theoretical values (dashed lines). Left: Experiment, right: simulation.

**Figure 8** BER vs. fiber launch power for transmission of 3D-Simplex (circles) in comparison to DP-DPSK (triangles). Upper: experiment, lower: simulation.

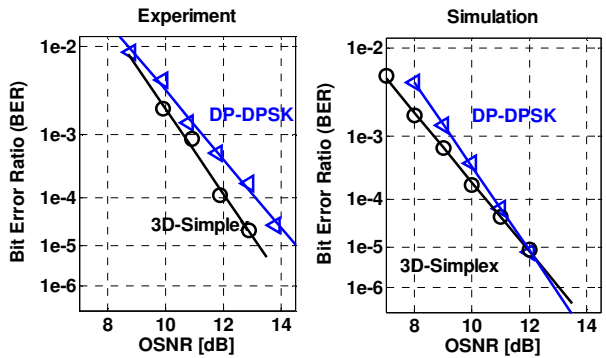

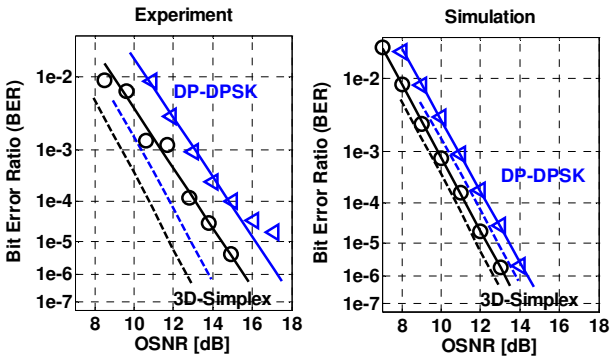

**Figure 7** Back-to-back BER results for 25 GBaud. 3D-Simplex (circles) and DP-DPSK (triangles) with solid lines as linear regression in comparison to theoretical values (dashed lines). Left: Experiment, right: simulation.

**Figure 9** BER vs. OSNR after 300 km transmission at optimum launch power (DP-DPSK: 17 dBm (experiment)/16 dBm (simulation), 3D-Simplex:16 dBm), increasing span loss (and thus reducing OSNR).

From both, measurement and simulation, it is visible that 3D-Simplex exhibits 1 dB less nonlinear tolerance than DP-DPSK. Observing the phase difference between BPSK and QPSK tributary of the 3D-Simplex signal after nonlinear transmission, the QPSK part experiences stronger non-linear phase rotation. This is partly compensated by the phase difference estimation (see section 3.3). However, it still leads to the slightly worse performance compared to DP-DPSK. The simulation results for this scenario indicate slightly smaller BER values, which might result since non-linear phase noise was not included in the numerical system setup. The wider opening of the simulative curves results from the better convergence of the standard CMA for a distorted signal in comparison to the methods used in the experiments. However, since the real system is never as stable as the simulation, it is im-possible to use a standard CMA there, as explained in [10].

In a next step, the optimum fiber launch power for each format was chosen to evaluate the resilience towards increased span loss. Within the experiments, the optimum input power of 17 dBm for DP-DPSK is 1 dB higher than for 3D-Simplex. The simulations indicate the same optimum input power of 16 dBm for both formats.

Setting the fiber launch power to the optimum value, the attenuation after 100 km fiber (see VOA in **Fig. 2**) is stepwise increased. The measured OSNR for the lowest attenuator loss of 0 dB corresponds to an OSNR of 13.9 dB for DP-DPSK at 17 dBm launch power and of 12.9 dB for 3D-Simplex at 16 dBm launch power, while an OSNR of 9.9 dB is obtained for an attenuation of 4 dB in case of DP-DPSK and for 3 dB in case of 3D-Simplex. Within the simulations the noise loading is performed at the receiver after transmission and the OSNR is adjusted according to the measured values. Experimental and simulation results are presented in **Fig. 9**.

From both, simulations and experiments, it can be seen that in the noise dominated transmission (left part of the curves in **Fig. 8** and low OSNR in **Fig. 9**) 3D-Simplex performs better than DP-DPSK. However, if the fiber input power exceeds 16 dBm DP-DPSK starts to outperform 3D-Simplex, allowing about 1 dB more input power at the same BER. The simulative curve of **Fig. 9** illustrates the case where DP-DPSK nearly equals 3D-Simplex. This curve is not exactly comparable to the experimental curve, since for the experiments the DP-DPSK power is higher by 1 dB compared to 3D-Simplex, thus worsening the DP-DPSK results.

## 5  Conclusion

For a 3-dimensional Simplex modulation format which encodes two bits per symbol, we demonstrated a performance gain of approximately 1 dB in comparison to DP-DPSK by means of experiments and Monte-Carlo simulations. Transmission over a 300 km link with varying channel launch power showed that up to an optimum power, which was identical for both formats, 3D-Simplex outperforms DP-DPSK. At higher-than-optimal channel launch values, DP-DPSK allows approximately 1 dB more input power as 3D-Simpex at the same BER. Experimental results are in good agreement with simulation.